\documentclass{article}

\newcommand{\p}[1]{(\ref{#1})}

\newcommand{\cF}{{\cal F}}

\newcommand{\bD}{{\overline D}{}}

\newcommand{\eps}{\varepsilon}

\newcommand{\be}{\begin{equation}}
\newcommand{\ee}{\end{equation}}
\newcommand{\bea}{\begin{eqnarray}}
\newcommand{\eea}{\end{eqnarray}}

\newcommand{\ba}{\begin{array}} \newcommand{\ea}{\end{array}}

\newcommand{\nn}{\nonumber}

\usepackage{amscd,amsmath,amssymb}

\begin{document}
\thispagestyle{empty}
\vspace{2cm}
\begin{flushright}
\end{flushright}\vspace{2cm}
\begin{center}
{\Large\bf N=4 supersymmetric 3-particles Calogero model}
\end{center}
\vspace{1cm}

\begin{center}
{\large\bf S.~Bellucci${}^{a}$, S.~Krivonos${}^{b}$,
A.~Sutulin${}^{b}$ }
\end{center}

\begin{center}
${}^a$ {\it
INFN-Laboratori Nazionali di Frascati,
Via E. Fermi 40, 00044 Frascati, Italy} \vspace{0.2cm}

${}^b$ {\it
Bogoliubov  Laboratory of Theoretical Physics, JINR,
141980 Dubna, Russia} \vspace{0.2cm}

\end{center}
\vspace{2cm}

\begin{abstract}\noindent
We constructed the most general $N=4$ superconformal 3-particles systems with translation invariance.
In the basis with decoupled center of mass the supercharges and Hamiltonian possess one arbitrary function which
defines  all potential terms. We have shown that with the proper choice of this function one may describe the
standard, $A_2$ Calogero model as well as $BC_2, B_2,C_2$ and $D_2$ Calogero models with $N=4$ superconformal symmetry.
The main property of all these systems is that even with the coupling constant equal to zero they still contain
nontrivial interactions in the fermionic sector. In other words, there are infinitely many non equivalent
$N=4$ supersymmetric extensions of the free action depending on one arbitrary function.
We also considered  quantization and explicitly showed how the supercharges and Hamiltonian are modified. 
\end{abstract}

\newpage
\setcounter{page}{1}

\section{Introduction}
The famous action of the $n$-particles Calogero model reads \cite{cal_gen}
\be\label{bac1}
S_{(n)}=\int dt \left[ \frac{1}{2} \sum_{i=1}^n{\dot x}{}_i^2 -\sum_{i<j} \frac{2g}{(x_i-x_j)^2}\right],
\ee
where the $n$ coordinates $x_i$ depend on the time $t$ only. This action describes a system of $n$ identical
particles on the line with pairwise interactions. All  models with  actions \p{bac1}
are invariant under conformal transformations in one-dimension. The standard description of this
invariance consists in the statement that the Hamiltonian for the action \p{bac1}
\be\label{bacham1}
H= \frac{1}{2} \sum_{i=1}^n{p_i}^2 +\sum_{i<j} \frac{2g}{(x_i-x_j)^2},
\ee
forms the $so(1,2)$ algebra together with the generators of the dilatation $D$ and  conformal boost $K$ defined
as
\be\label{bacDK1}
D= \sum_{i=1}^n x_i p_i, \qquad K= \sum_{i=1}^n x_i{}^2,
\ee
with respect to canonical Poisson brackets
\be\label{bacPB}
\left\{ x_i, p_j\right\} = \delta_{ij}.
\ee
In this paper we consider a $N=4$ supersymmetric generalization of the particular case of the system \p{bac1} with three particles. 
As its basic property, the constructed system will possess invariance with respect to the $N=4$
super-extension of the $so(1,2)$ symmetry -- i.e. the  $su(1,1|2)$ superalgebra.

The case of the 3-particles Calogero model is the first nontrivial one. Indeed, in the case when  $n=1$, the action \p{bac1}
is  just a free action for the center of mass $X_0=x_1$
\be\label{bacn1}
S_{(1)}=\int dt\, \frac{1}{2} \dot{X}_0{}^2 .
\ee
The next case with $n=2$ is described by the action
\be\label{bac2}
S_{(2)}=\int dt \left[ \frac{1}{2}\left( {\dot x}_1{}^2+{\dot x}_2{}^2\right)  -\frac{2g}{(x_1-x_2)^2}\right].
\ee
After passing to the coordinates
\be
X_0=\frac{1}{\sqrt{2}}\left(x_1+x_2\right), \qquad z=\frac{1}{\sqrt{2}}\left(x_1-x_2\right)
\ee
it acquires the following form:
\be\label{bac2a}
S_{(2)}=\int dt \left[ \frac{1}{2}\left( {\dot X}_0{}^2+{\dot z}^2\right)  -\frac{g}{z^2}\right].
\ee
So, in this case we have a direct sum of two actions: one again describes the free motion
of the center of mass of the system with the coordinate $X_0$, while the second one is the action of conformal mechanics \cite{FF}.
The $N=4$ supersymmetrization of these two cases is rather simple. In the case with $n=1$, we need just the free
$N=4$ supersymmetric action for the particle in one dimension, while for the two particles case the corresponding $N=4$ action is a direct sum of 
the $N=4$ free action and the action of $N=4$ superconformal mechanics \cite{leva}. It is worth noting that, for $n=1,2$, the actions are completely
fixed if one insists on the conformal invariance.

The main new feature which appears for the $n\geq 3$ cases is the fact that the conformal invariance is
not enough to completely fix the action. Indeed, even for the 3-particles case one may construct
infinitely many conformally invariant actions (see the discussion at the end of the next Section). In this respect, the Calogero
models should be picked up from this huge set of conformally invariant models by some additional, still unknown
properties or symmetries. In addition, the $N=4$ superconformal symmetry puts strong conditions on the possible potential terms.
It was shown in \cite{leva} that the proper potential for the one-particle $N=4$ superconformal mechanics can not be added by hand, instead 
it is generated automatically from the superfield $\sigma$-model terms due to the presence of a constant among
the auxiliary components of the corresponding $N=4$ supermultiplet. Thus, the structure of the potential
cannot be freely generated as in case of $N=2$ supersymmetry \cite{freedman}.
All these features forced us to consider firstly the simplest case of the $N=4$ supersymmetric extension of the 3-particles Calogero models, which is the
main subject of the present paper.

The first attempt to construct the $N=4$ supersymmetric extension of the Calogero models was  performed
by N.~Wyllard in \cite{Wyllard}. The result was completely discouraging. Indeed, it was shown that such a system does not exist at all. 
The next important step has been done recently in \cite{GLP} where the supercharges and Hamiltonian were explicitly constructed for the $N=4$ 
supersymmetric 3-particles Calogero model. In this paper the Authors showed
that the terms which spoil the  construction by Wyllard can be interpreted as  quantum corrections. So, in the classical limit the
proper action \p{bac1} could be obtained.
Unfortunately, the component description in the Hamiltonian formalism presented in \cite{GLP} being extended to $n > 3$ particles 
cases leads to a very complicated systems of equations for which even the proof of existence of solutions is rather nontrivial. 

In the present paper we show that the $N=4$ supersymmetric 3-particles Calogero model has a very natural formulation in terms of $N=4$ superfields. 
Moreover, we constructed the most general $N=4$ superconformally invariant 3-particles action which contains, as particular cases,  
the known $A_2,G_2$ and $BC_2$ Calogero models. These results have been achieved by excluding the center of mass and passing to a new set of coordinates.  
In addition, we explicitly demonstrated that there are infinitely many $N=4$ supersymmetric extensions of
the free 3-particles system and only one of these extensions could be lifted up to the $N=4$ Calogero model.

\setcounter{equation}0
\section{Preliminaries: the bosonic case}
The bosonic action of the 3-particles Calogero model
\be\label{bac3}
S_{(3)}=\int dt \left[ \frac{1}{2} \sum_{i=1}^3{\dot
x}{}_i^2 -\sum_{i<j} \frac{2g}{(x_i-x_j)^2}\right],
\ee
is not too convenient for supersymmetrization. First of all, after passing to the new coordinates
\be\label{coor1}
X_0=\frac{1}{\sqrt{3}}\left(x_1+x_2+x_2\right), \quad
y_1=\frac{1}{\sqrt{6}}\left(2x_1-x_2-x_3\right),\;
y_2=\frac{1}{\sqrt{2}}\left(x_2-x_3\right),
\ee
one may decouple the motion of the center of mass, which is described by the coordinate $X_0$,  from the rest of the coordinates
\be\label{bac3a}
S_{(3)}=S_0+S=\int dt \left\{ \frac{1}{2}{\dot
X}_0^2 + \frac{1}{2}\sum_{i=1}^2 {\dot y}{}_i^2 -g\left[
\frac{1}{y_2^2}+\frac{4}{\left(\sqrt{3} y_1+y_2\right)^2}
+\frac{4}{\left(\sqrt{3} y_1-y_2\right)^2}\right] \right\}.
\ee
Clearly, the supercharges for such a system will be just the sum of two sets of commuting supercharges -
one set for a "free particle" $X_0$ and the second one for the interacting particles $y_1,y_2$.
The supersymmetrization of the free action $S_0$ goes straightforwardly, so we will restrict our
attention in this paper to the supersymmetrization of the action $S$ \p{bac3a}. Even in this case, one may further
simplify the action by passing to the new coordinates
\be\label{coor2}
y_1=e^u \sin{\phi},\quad y_2=e^u \cos{\phi}.
\ee
In these coordinates the action $S$ \p{bac3a} acquires the following  form:
\be\label{bac3b}
S=\int dt\left[ \frac{1}{2} e^{2u} \left( {\dot u}{}^2 +{\dot\phi}{}^2\right)
-9g\;\frac{e^{-2u}}{\cos^2{(3\phi})}\right].
\ee
So, our task in this paper is to construct the $N=4$ superconformal extension of this action.

Before closing this Section it makes sense to comment on the conformal invariance of the action \p{bac3b}.
In one dimension the conformal group is $SO(1,2)$, and it has the following realization on the time $t$ {}\footnote{In
accordance with the
standard definition of the conformal transformations in $d=1$ they are just general coordinate transformations
$\delta t = f(t)$ without any restrictions on the function $f$. The finite dimensional subgroup of this
infinite dimensional  group is $SO(1,2)$ which is selected by the constraint \p{conf1} on the function $f(t)$.}:
\be\label{conf1}
\delta t = f(t), \qquad \frac{d^3}{dt^3} f(t)=0 \; \Rightarrow \; f(t)=a+bt+ct^2.
\ee
The parameters $(a,b,c)$ correspond to the translations, dilatations and conformal boosts,
respectively. In contrast with the standard treatment of the conformal invariance of the Calogero model we
shortly discussed in the Introduction,
one may check that the action \p{bac3b} is explicitly invariant under \p{conf1} provided the
field $u$ transforms as a dilaton
\be\label{conf2}
\delta u = \dot{f}(t),
\ee
while $\phi$ is a scalar under conformal transformations. In this paper we will adhere just to this formulation
of the conformal invariance and will extend it to the $N=4$ supersymmetric case in the next Sections.

Finally, with our interpretation of the conformal invariance, it becomes almost evident that it does not fix the action in the 
3-particles case. 
Indeed, the most general conformally invariant action for the
dilaton $u(t)$ and arbitrary scalar field $\phi(t)$ reads
\be\label{conff}
{\tilde S}\sim \int dt \left[\frac{1}{2} e^{2u} \left( {\dot u}{}^2 +{\dot\phi}{}^2\right)- g e^{-2u} V(\phi) \right],
\ee
where $V(\phi)$ is an arbitrary function of its argument and $g$ is a coupling constant. Clearly, the Calogero action \p{bac3b} is a very 
special case of the general conformally invariant action \p{conff}. One has to mention, that
the standard treatment of the conformal invariance of the Calogero models leads to the same conclusion.
Indeed, in the new variables $x= e^u$ the Hamiltonian for the action \p{conff} reads
\be\label{haM}
H= \frac{ p_x{}^2}{2} +\frac{1}{x^2}\left[ \frac{{p_\phi}^2}{2}+g V(\phi)\right].
\ee
It is a rather evident that the Hamiltonian \p{haM} forms the $so(1,2)$ algebra together with the following
generators of the dilatations and conformal boosts:
\be
D=x p_x, \qquad K=x^2,
\ee
similarly to \p{bacham1}, \p{bacDK1}. Clearly, the potential $V(\phi)$ is still a completely arbitrary function and,
in order to bring it to 
Calogero form \p{bac3b}, one has to impose additional requirements besides the conformal invariance.

\setcounter{equation}0
\section{Superconformal invariance and superspace action}
In order to construct the $N=4$ supersymmetric extension of the action \p{conff}, we have to choose the proper $N=4$ supermultiplets
which will contain our bosonic fields $u(t)$ and $\phi(t)$ and then construct the superconformally invariant
action which will be reduced to \p{conff} in the  bosonic limit. Concerning the suitable supermultiplets we do not have
too much freedom. It has been known for a long time that there is only one $N=4$ supermultiplet with one physical
boson among its components \cite{leva}. So, we will need two such multiplets. In the $N=4, d=1$ superspace
with the coordinates $(t, \theta_i,\bar\theta{}^i)$ they are described by a pair of real bosonic superfields
$Y, V$ subjected to the following constraints \cite{leva}:
\be\label{sf1}
\left[ D^i, \bD_i \right]Y=0, \qquad \left[ D^i, \bD_i \right]V=0,
\ee
where the covariant spinor derivatives obey the following algebra:
\be\label{der}
\left\{ D^i, \bD_j \right\}=2i\delta^i_j \partial_t, \quad \left\{ D^i, D^j \right\}=\left\{ \bD_i, \bD_j \right\}=0.
\ee
It is worth to note that immediate consequences of the constraints \p{sf1} are
\be\label{sf2}
\left\{
\begin{array}{l}
\frac{d}{dt} D^2 Y=\frac{d}{dt} D^2 V=0,\\
\\
 \frac{d}{dt} \bD{}^2 Y=\frac{d}{dt} \bD{}^2 V=0
\end{array}
\right. \qquad \Rightarrow \quad
\left\{
\begin{array}{l}
D^2 Y=i M, \quad D^2 V=im,\\
\\
\bD{}^2 Y=-i M, \quad \bD{}^2 V=-i m,
\end{array}
\right.
\ee
where $m$ and $M$ are constants. So, among the auxiliary components in the superfields $Y$ and $V$ there are
some arbitrary constants. Just these constants will give rise to a potential term in the full action \cite{leva}.

The next step is to construct the superconformally invariant action in terms of our $N=4$ superfields
$Y,V$. In one dimension the most general $N=4$ superconformal group is $D(2,1;\alpha)$  \cite{sorba}.
Here we restrict our consideration to the special case of $D(2,1;\alpha)$ with $\alpha=-1$, which corresponds to the $SU(1,1|2)$
superconformal group. This group may be naturally realized in $N=4, d=1$ superspace as \cite{leva}
\be\label{sc1}
\delta t = E -\frac{1}{2} \theta_i D^i E -\frac{1}{2}\bar\theta{}^i \bD_i E, \quad
\delta \theta_i=-\frac{i}{2} \bD{}_i E,\quad
\delta \bar\theta{}^i=-\frac{i}{2} D^i E,
\ee
where the superfunction $E(t,\theta,\bar\theta)$ collects all $SU(1,1|2)$ parameters
\be\label{E}
E=f(t)-2i \left( \varepsilon \bar\theta -\theta\bar\varepsilon \right) + \theta^i \bar\theta{}^j B_{(ij)}+
  2\left( \dot\varepsilon \bar\theta +\theta\dot{\bar\varepsilon}\right)(\theta\bar\theta) +\frac{1}{2} (\theta\bar\theta)^2
   \ddot{f}.
\ee
Here
\be
f=a+bt+ct^2,\quad \varepsilon^i = \epsilon^i + t \eta^i.
\ee
The bosonic parameters $a,b,c$ and $B_{(ij)}$ correspond to translations, dilatations, conformal boosts and
rigid $SU(2)$ rotations, while the fermionic parameters $\epsilon^i$ and $\eta^i$ correspond to Poincar\`e and
conformal supersymmetries, respectively.
One may check  that by construction the function $E$ \p{E}  obeys the conditions
\be\label{cE}
D^2 E = \bD{}^2E= \left[D^i,\bD_i\right]E=0, \qquad \partial^3_t E = \partial^2_t D^i E =
\partial_t D^{(i}\bD{}^{j)}E=0.
\ee
It is a rather important that the spinor derivatives \p{der}  transform  under $SU(1,1|2)$  as \cite{ikl1}
\be\label{trD}
\delta D^i =-\frac{i}{2} \left( D^i \bD_j E\right)D^j, \qquad \delta \bD_i =
-\frac{i}{2}\left( \bD_i D^j E\right) \bD_j.
\ee
One may check that, in view of \p{trD}, the constraints \p{sf1} are invariant under the $N=4$ superconformal group
$SU(1,1|2)$ only if the superfields $Y, V$ nontrivially transform as
\be\label{sc2}
\delta Y = \partial_t E\; Y, \qquad \delta V=\partial_t E\; V.
\ee
So, the superfields $Y, V$ are vectors under superconformal transformations.

In order to construct a superconformally invariant action, one has take into account that the superspace measure $ds$
\be\label{sc22}
ds = dt d^2 \theta d^2 \bar\theta ,
\ee
is also transformed under \p{sc1} as
\be\label{measure}
\delta ds = - \partial_t E \; ds.
\ee
Thus, the superconformally invariant measure can be constructed in the two different ways
\be\label{maesure1}
\triangle_1 s = Y ds, \qquad \mbox{  or } \qquad \triangle_2 s = V ds.
\ee
With all these ingredients we are ready to write the $N=4$ superconformally invariant action for the one
particle case \cite{leva}
\be\label{n4n1}
S_{(1)}=-\frac{1}{2}\int\; \triangle_1 s \; \log Y \equiv -\frac{1}{2}\int\; ds \; Y \log Y .
\ee
In order to check the invariance of this action, it is crucial to use the properties \p{cE}. Indeed, the variation of the
action \p{n4n1} reads
\be\label{var}
\delta S_{(1)}=-\frac{1}{2}\int\; ds \partial_t E \;Y .
\ee
All terms in the r.h.s. of this variation disappear after integration over Grassmann coordinates only in virtue
of the constraints \p{cE}.

The bosonic sector of the action \p{n4n1} reads\footnote{We defined the superspace
measure as $d^2 \theta d^2 \bar\theta = \frac{1}{16} D^2 {\bar D}^2$.}
\be\label{n4n1bos}
S_{(1)}^{bos}=\frac{1}{32}\int dt\left[ 4 \frac{{\dot y}{}^2}{y} -\frac{M^2}{y}\right],
\ee
where $y$ is the first bosonic component of the $N=4$ superfield $Y$.
Clearly, after passing to the new variable $z=\sqrt{y}$ the action \p{n4n1bos} will coincide with \p{bac2a} (without center of mass $X_0$)
upon the identification $g=\frac{M^2}{32}$. This is the well known case of the $N=4$ supersymmetric one particle Calogero
model. It is evident that the two-particles case can be easily constructed as a direct sum of  two  such actions.
The main lesson we have learned  from these simplest cases is how to automatically generate  the corresponding
potential terms thanks to the presence of  constants among the components of our superfields. This makes all
the $N=4$ supersymmetric  construction strictly rigid.

The general bosonic action for the 3-particles case in the basis where the center of masses is decoupled \p{conff}
contains an arbitrary function $V(\phi)$. So, for extending the above construction to the 3-particles case, i.e. the case with
two superfields $Y$ and $V$, one has firstly  to construct a scalar superfield from $Y$ and $V$ as follows:
\be\label{Z}
Z=V Y^{-1}\qquad \Rightarrow \quad\delta Z=0.
\ee
Then, the general superconformally invariant action constructed from the superfield $Y$ and
scalar superfield $Z$ has the form
\be\label{Action}
S_{(3)}^{gen}=-\frac{1}{2}\int ds \; Y \left[  \log{Y}+ F\left( Z \right) \right],
\ee
where  $F\left( Z \right)$ is an arbitrary function of the scalar superfield $Z$ \p{Z}. 
The invariance of the first term in the action \p{Action} has been already
demonstrated above, while the second term is manifestly invariant by construction. 

Our  main statement is that the $N=4$ supersymmetric 3-particles Calogero model lies inside the class of the
general superconformally invariant actions \p{Action}. In order to prove this, let us firstly consider the bosonic part of the
component action which follows from \p{Action} after integrating over Grassmann variables and putting all fermions to zero.
This bosonic action reads
\be\label{ba1}
S_{(3)}^{bos}=\frac{1}{32}\int dt \left[ 4\frac{{\dot y}{}^2}{y}+\frac{M^2}{y} + F_{zz}\left( 4 y {\dot z}{}^2 - 
\frac{(m-Mz)^2}{y}\right) \right].
\ee
Here $z$ is the first bosonic component of the $N=4$ superfield $Z$. Now we put $M=0$, choose 
\be\label{F}
F_{zz}=4\left( \phi_z \right)^2
\ee
and pass to the new coordinate $u=\frac{1}{2}\log(y)$
\be\label{ba2}
S_{(3)}^{bos}=\int dt \left[ \frac{1}{2}e^{2u}\left( {\dot u}{}^2+{\dot\phi}{}^2\right) - \frac{e^{-2u}m^2}{8} \left(\phi_z \right)^2\right].
\ee
Here we used the equality $\dot\phi=\phi_z {\dot z}$ to bring the kinetic term for  $\phi$ to the standard form. It is clear now that if we will
choose
\be\label{eqphi}
\left(\phi_z \right)^2 = \frac{1}{9 \cos^2(3\phi)},
\ee
then the action \p{ba2} will coincide with \p{bac3b} upon the identification $9g=m^2/72$. The last step is to reconstruct the function $F(z)$ from
\p{F} and \p{eqphi}
\be\label{FF}
F_{zz}=\frac{4}{9(1-z^2)} \quad \Rightarrow F=\frac{2}{9}\left[ (1+z) \log(1+z) +(1-z)\log(1-z)\right].
\ee
Thus, the bosonic sector of the $N=4$ superconformal action \p{Action} with the function $F$ defined in \p{FF} coincides with the action
of the 3-particles Calogero model \p{bac3b}. In the next Section we will find the full component form of this action.

\setcounter{equation}0
\section{N=4 supersymmetric 3-particles Calogero model}
With the function $F$ specified for the case of Calogero model in \p{FF}, the action \p{Action} may be rewritten as
\be\label{ActionS}
S_{(3)}=- \frac{1}{9}\int ds \left[ \frac{5}{2} Y \log{Y}+ \left( Y+V\right) \log{\left(Y+V\right)}+
\left( Y-V\right) \log{\left(Y-V\right)} \right].
\ee
\subsection{Component action}
In order to pass to components, one has to integrate over Grassmann variables. Before doing this, let us define the
component fields. The constraints \p{sf1} with $M=0$ leave the following components in the $N=4$ superfields $V$ and $Y$:
\bea
&&
v = \left. V \right|\,, \;
\psi^i = \left. i D^i V \right|\,, \;
\bar \psi_i = \left. i \bar D_i V \right|\,, \;
A^i_j = \frac{1}{2} [D^i, \bar D_j]  \left. V \right|\,, \;
i m  = \left. D^2 V \right|\,, \;
- i m =  \left. \bar D^2 V \right|\,,
\nn\\
&&
y = \left. Y \right|\,, \quad
\lambda^i = \left. i D^i Y \right|\,, \quad
\bar \lambda_i = \left. i \bar D_i Y \right|\,, \quad
B^i_j =  \frac{1}{2} [D^i, \bar D_j] \left. Y \right|\,,  \nn\\
\label{Z1}
\eea
where, as usual, $|$ means restriction to $\theta_i=\bar\theta{}^i=0$. 
Now one may integrate over $\theta's$ to get the full off-shell component action of the model
\bea
S \!&=&\! \frac{1}{144}\int dt \bigg \{
10 \frac{{\dot y}^2}{y}+4 \frac{\left({\dot y}+{\dot v}\right)^2}{y+v}+
4 \frac{\left({\dot y}-{\dot v}\right)^2}{y-v}
\nn\\
\!&+&\! 2i \frac{9y^2-5 v^2}{y(y^2- v^2)}
[\bar \lambda^i \dot {\lambda}_i - \lambda^i \dot{\bar \lambda}_i]
+ 8i \frac{y}{y^2- v^2}
[\bar \psi^i \dot \psi_i - \psi^i \dot{\bar \psi}_i]
+ 8i \frac{u}{y^2- v^2}
[\lambda^i \dot{\bar \psi}_i - \bar \psi^i {\dot \lambda}_i
+ \psi^i \dot{\bar \lambda}_i - \bar \lambda^i {\dot \xi}_i]
\nn\\
\!&-&\! 2 m^2 \frac{y}{y^2-v^2}
- 4i m \frac{v y}{(y + v)^2 (y - v)^2}
[\lambda^2 + \psi^2 - \bar \lambda^2 - \bar \psi^2]
+ 4i m \frac{y^2+v^2}{(y + v)^2 (y - v)^2}
[\lambda^i \psi_i - \bar \lambda_i \bar \psi^i]
\nn\\
\!&-&\!
\frac{5}{y} B^{(ij)} B_{(ij)}
- \frac{2}{y+v} (B^{(ij)} + A^{(ij)}) (B_{(ij)} + A_{(ij)})
- \frac{2}{y-v} (B^{(ij)} - A^{(ij)}) (B_{(ij)} - A_{(ij)})
\nn\\
\!&-&\! \frac{10}{y^2} \lambda_{(i} {\bar \lambda}_{j)}B^{(ij)}
- \frac{4}{(y+v)^2} (\lambda_i + \psi_i) (\bar \lambda_j + \bar \psi_j) (B^{(ij)} + A^{(ij)})
- \frac{4}{(y-v)^2} (\lambda_i - \psi_i) (\bar \lambda_j - \bar \psi_j) (B^{(ij)} - A^{(ij)})
\nn\\
\!&-&\! \frac{5}{y^3} \lambda^2 \bar \lambda^2
- \frac{2}{(y+v)^3} (\lambda^i + \psi^i) (\lambda_i + \psi_i) (\bar \lambda_j + \bar \psi_j) (\bar \lambda^j + \bar \psi^j)
- \frac{2}{(y-v)^3} (\lambda^i - \psi^i) (\lambda_i - \psi_i) (\bar \lambda_j - \bar \psi_j) (\bar \lambda^j - \bar \psi^j)
\bigg \}
\nn\\
\label{Z2}
\eea
The $N=4$ supersymmetry transformations
$\delta \Psi = i(\bar \eps_i Q^i + \eps^i \bar Q_i) \Psi$
are realized on the components \p{Z1} as
\bea
&&
\delta v = \eps^i \bar \psi_i + \bar \eps_i \psi^i\,, \quad
\delta \psi^i = - i \eps^i \dot v - \eps_k A^{(ik)} - \frac{i}{2} \bar \eps^i m\,, \quad
\delta \bar \psi_i = \frac{i}{2} \eps_i m - i \bar \eps_i \dot v + \bar \eps^k A_{(ik)} \nn\\
&&
\delta y = \eps^i \bar \lambda_i + \bar \eps_i \lambda^i\,, \quad
\delta \lambda^i = -i \eps^i \dot y - \eps_k B^{(ik)}\,, \quad
\delta \bar \lambda_i = -i \bar \eps_i \dot y + \bar \eps^k B_{(ik)}\,.
\label{Z3}
\eea
One may check that the action \p{Z2} is invariant with respect to these transformations.

In order to bring the action \p{Z2} to the familiar form (at least in the bosonic sector) one has to introduce new physical bosonic fields
\be
y = e^{2u}\,, \qquad v = e^{2u} \sin(3\phi).
\label{Z4}
\ee
For simplifying  the component action, we also introduce a new set of spinor variables $(\eta, \xi)$ which are related to the old ones as
\bea
&&
\psi^i = 3 e^u \cos (3\phi)\, \xi^i + 2 e^u \sin (3\phi)\, \eta^i\,, \quad
\lambda^i = 2 e^u \eta^i\,, \nn\\
&&
\bar \psi_i = 3 e^u \cos (3\phi)\, \bar \xi_i + 2 e^u \sin (3\phi)\, \bar \eta_i\,, \quad
\bar \lambda_i = 2 e^u \bar \eta_i\,.
\label{Z6}
\eea
The next step is to use the equations of motion for auxiliary fields to express them in terms of physical components 
\bea
&&
B_{(ik)} = -4 \xi_{(i} \bar \xi_{k)} -4 \eta_{(i} \bar \eta_{k)}\,, \nn\\
&&
A_{(ik)} = 14 \sin(3\phi) \xi_{(i} \bar \xi_{k)} -4 \sin(3\phi) \eta_{(i} \bar \eta_{k)}
- 6 \cos(3\phi) \xi_{(i} \bar \eta_{k)} - 6 \cos(3\phi) \eta_{(i} \bar \xi_{k)}\,.
\label{Z7}
\eea
Finally, using (\ref{Z7}) one may find the on-shell action of $N=4$ 3-particles Calogero model
\bea
S \!&=&\! \int dt \bigg \{ \frac{1}{2} e^{2u} \left({\dot u}^2+ {\dot \phi}^2\right)
+ \frac{i}{2} (\bar \eta^i \dot{\eta}_i - \eta^i \dot{\bar \eta}_i)
+ \frac{i}{2} (\bar \xi^i \dot{\xi}_i - \xi^i \dot{\bar \xi}_i)
+ i \dot \phi (\eta^i \bar \xi_i - \xi^i \bar \eta_i) \nn\\
\!&-&\! m^2 e^{-2u} \; \frac{1}{72\cos^2({3\phi})}
- i m e^{-2u} \frac{\sin (3\phi)}{4\cos^2 (3\phi)} (\xi^2 - \bar \xi^2)
+ i m e^{-2u} \frac{1}{6\cos (3\phi)} (\xi^i \eta_i - \bar \xi_i \bar \eta^i) \nn\\
\!&-&\! \frac{1}{4} e^{-2u} \left (\eta^2 \bar \eta^2 - 3 \left(1 - \frac{3}{\cos^2 (3\phi)} \right)
\xi^2 \bar \xi^2 \right)
- \frac{1}{4} e^{-2u} \left (\eta^2 \bar \xi^2 + \xi^2 \bar \eta^2 \right) \nn\\
\!&-&\! e^{-2u} \left ( \eta^i \bar \eta_i \xi^j \bar \xi_j +  \eta^i \bar \xi_i \xi^j \bar \eta_j \right)
+ \frac{3}{2} e^{-2u} \tan (3\phi) \left(\eta^i \xi_i \bar \xi^2 + \xi^2  \bar \eta_i \bar \xi^i \right)
\bigg \}
\label{Z8}
\eea
\subsection{Supercharges and Hamiltonian}
The explicit form of the on-shell action \p{Z8} together with the $N=4$ supersymmetry transformations \p{Z3}, \p{Z7} provide
all ingredients needed to construct the supercharges and Hamiltonian.

First of all, we define the  momenta for bosonic and fermionic components
\bea
&& P_u = e^{2u} \dot u\,, \qquad
P_{\phi} = e^{2u} \dot \phi + i (\eta^i \bar \xi_i - \xi^i \bar \eta_i), \label{Z9} \\
&& \pi_\xi^i=\frac{i}{2} {\bar\xi}{}^i, \qquad \pi_\eta^i=\frac{i}{2} {\bar\eta}{}^i. \label{Z9a}
\eea
As usual, the fermionic momenta \p{Z9a} mean that the system possesses  second class constraints. Therefore, one has pass
to  Dirac brackets for the canonical variables
\be
\{u, P_u \}_D = 1\,, \quad
\{\phi, P_{\phi} \}_D = 1\,, \quad
\{\eta^i, \bar \eta_j \}_D = i \delta^i_j\,, \quad
\{\xi^i, \bar \xi_j \}_D = i \delta^i_j\,. \quad
\label{Z10}
\ee
Now one may check that the supercharges $Q^i, {\bar Q}_i$
\bea
&&
Q^i = e^{-u} \eta^i P_u +  e^{-u} \xi^i P_{\phi}
- \frac{m}{6} \frac{e^{-u}}{\cos(3\phi)} \bar \xi^i  - \frac{3i}{2} e^{-u} \tan(3\phi) \xi^2 \bar \xi^i
+ \frac{i}{2} e^{-u} \eta^2 \bar \eta^i - 2i e^{-u} \xi^{(i} \bar \xi^{k)} \eta_k\,, \nn\\
&&
{\bar Q}_i =  e^{-u} \bar \eta_i P_u +  e^{-u} \bar \xi_i P_{\phi}
+ \frac{m}{6} \frac{e^{-u}}{\cos(3\phi)} \xi_i  - \frac{3i}{2} e^{-u} \tan(3\phi) \xi_i \bar \xi^2
+ \frac{i}{2} e^{-u} \eta_i \bar \eta^2 + 2i e^{-u} \xi_{(i} \bar \xi_{k)} \bar \eta^k\,. \nn\\
\label{Z11}
\eea
and the Hamiltonian
\bea
H \!&=&\! \frac{1}{2} e^{-2u} P^2_u +
\frac{1}{2} e^{-2u} P^2_{\phi}
- i e^{-2u} P_{\phi} \left (\eta^i \bar \xi_i - \xi^i \bar \eta_i \right ) \nn\\
\!&+&\! \frac{1}{72} e^{-2u} m^2 \frac{1}{\cos^2 (3\phi)}
- \frac{i}{6} e^{-2u} m  \left (\frac{1}{\cos (3\phi)} \eta^i  \xi_i
-\frac{1}{\cos (3\phi)} \bar \eta_i \bar \xi_i
- \frac{3}{2} \frac{\sin (3\phi)}{\cos^2 (3\phi)} \xi^2 + \frac{3}{2} \frac{\sin (3\phi)}{\cos^2 (3\phi)} \bar \xi^2 \right) \nn\\
\!&+&\! \frac{1}{4} e^{-2u} \left (\eta^2 \bar \eta^2 - 3 \left(1 - \frac{3}{\cos^2 (3\phi)} \right)
\xi^2 \bar \xi^2 \right) \nn\\
\!&+&\! e^{-2u} \left ( \eta^i \bar \eta_i \xi^j \bar \xi_j + 2 \eta^i \bar \xi_i \xi^j \bar \eta_j \right)
- \frac{3}{2} e^{-2u} \tan (3\phi) \left(\eta^i \xi_i \bar \xi^2 + \xi^2  \bar \eta_i \bar \xi^i \right)
\label{Z12}
\eea
form the $N=4$ superalgebra
\be\label{sca}
\left\{ Q^i , {\bar Q}_j\right\}=2i \delta^i_j H .
\ee
With this, we completed the classical description of the 3-particles Calogero model. 

\setcounter{equation}0
\section{N=4 supersymmetric systems: general 3-particles case, $G_2$, $BC_2, B_2,C_2$ and $D_2$  Calogero models}
In the previous Section we considered the particular case of the general $N=4$ superconformal action \p{Action}
for the most interesting case of the $A_2$ Calogero model corresponding to the superfunction $F(Z)$ specified in \p{FF}.
But the general case could be also treated in the same manner. Indeed, one may check that the supercharges 
\bea
&&
Q^i = e^{-u} \eta^i P_u +  e^{-u} \xi^i P_{\phi}
- \frac{m}{6} e^{-u} \cF(\phi) \bar \xi^i  - \frac{i}{2} e^{-u} \frac{\cF^{\prime}(\phi)}{\cF(\phi)} \xi^2 \bar \xi^i
+ \frac{i}{2} e^{-u} \eta^2 \bar \eta^i - 2i e^{-u} \xi^{(i} \bar \xi^{k)} \eta_k\,, \nn\\
&&
\bar Q_i =  e^{-u} \bar \eta_i P_u +  e^{-u} \bar \xi_i P_{\phi}
+ \frac{m}{6} e^{-u} \cF(\phi) \xi_i  - \frac{i}{2} e^{-u} \frac{\cF^{\prime}(\phi)}{\cF(\phi)} \xi_i \bar \xi^2
+ \frac{i}{2} e^{-u} \eta_i \bar \eta^2 + 2i e^{-u} \xi_{(i} \bar \xi_{k)} \bar \eta^k\,
\label{Z13}
\eea
form with the Hamiltonian
\bea
H \!&=&\! \frac{1}{2} e^{-2u} P^2_u +
\frac{1}{2} e^{-2u} P^2_{\phi}
- i e^{-2u} P_{\phi} \left (\eta^i \bar \xi_i - \xi^i \bar \eta_i \right ) \nn\\
\!&+&\! \frac{1}{72} e^{-2u} m^2 \cF^2(\phi)
- \frac{i}{6} e^{-2u} m  \left ( \cF(\phi) \eta^i  \xi_i
-\cF(\phi) \bar \eta_i \bar \xi_i
- \frac{1}{2} \cF^{\prime}(\phi) \xi^2 + \frac{1}{2} \cF^{\prime}(\phi) \bar \xi^2 \right) \nn\\
\!&+&\! \frac{1}{4} e^{-2u} \left (\eta^2 \bar \eta^2 - \left(3 +\frac{(\cF^{\prime}(\phi))^2}{\cF^2(\phi)}- 
\frac{\cF^{\prime \prime}(\phi)}{\cF(\phi)} \right)
\xi^2 \bar \xi^2 \right) \nn\\
\!&+&\! e^{-2u} \left ( \eta^i \bar \eta_i \xi^j \bar \xi_j + 2
\eta^i \bar \xi_i \xi^j \bar \eta_j \right) - \frac{1}{2} e^{-2u}
\frac{\cF^{\prime}(\phi)}{\cF(\phi)} \left(\eta^i \xi_i \bar \xi^2 +
\xi^2  \bar \eta_i \bar \xi^i \right) \label{Z14} \eea
the same $N=4$ superalgebra \p{sca} with respect to the same brackets \p{Z10}.
Here, the function $\cF(\phi)$ is implicitly related to the prepotential $F(Z)$ as
\be
F_{zz}\equiv 4\left( \phi_z\right)^2 = \frac{4}{9}\cF^2(\phi).
\ee
These supercharges and Hamiltonian provide the most general solution for all possible 3-particles (with translation invariance) and
2-particles systems with $N=4$ superconformal invariance. As particular commonly interesting cases, let us consider in details
the $BC_2$ and $G_2$ Calogero models.

\subsection{$G_2$ Calogero model}
The rational $G_2$ model was originally proposed in \cite{G2}. Its action
\be\label{g1} 
S_{\rm G_2}=\int dt \left[ \frac{1}{2} \sum_{i=1}^3{\dot
x}{}_i^2 -\sum_{i<j} \frac{2g}{(x_i-x_j)^2}
-\sum_{i<j;i,j \neq k} \frac{6f}{(x_i-x_j+2x_k)^2}\right],
\ee
describes a 3-particles system with pairwise and three-body interactions. It is worth mentioning that in the
limit $f\rightarrow 0$, the $G_2$ model becomes the Calogero one \p{bac3}. After passing to the coordinates \p{coor1},
the center of mass with the coordinate $X_0$ completely decouples
\bea\label{g2}
S_{\rm G_2}=S_0+S \!&=&\! \int dt \left \{ \frac{1}{2}{\dot
X}_0^2
+  \frac{1}{2}\sum_{i=1}^2 {\dot y}{}_i^2 - g\left[
\frac{1}{y_2^2}+\frac{4}{\left(\sqrt{3} y_1+y_2\right)^2}
+\frac{4}{\left(\sqrt{3} y_1-y_2\right)^2}\right] \right. \nn\\
\!&-&\! \left. f \left[\frac{1}{y_1^2}+\frac{4}{\left(\sqrt{3} y_2+y_1\right)^2}
+\frac{4}{\left(\sqrt{3} y_2-y_1\right)^2}\right]
\right \}.
\eea
Finally, after introducing the new coordinates \p{coor2} the action acquires a very nice form
\be\label{g3}
S_{\rm G_2}=\int dt\left \{ \frac{1}{2} e^{2u} \left( {\dot u}{}^2 +{\dot\phi}{}^2\right)
-9e^{-2u}\;\frac{g}{\cos^2(3\phi)}
-9e^{-2u}\;\frac{f}{\sin^2(3\phi)}\right \}.
\ee
So, our task is to construct the $N=4$ superconformal extension of this action. Just comparing the Hamiltonian
for bosonic $G_2$ model with the general one in \p{Z14}, one may immediately fix the function ${\cal F}$ to read
\be\label{F1}
\frac{1}{72} m^2 \cF^2(\phi) = 9\left( \frac{g}{\cos^2(3\phi)}+\frac{f}{\sin^2(3\phi)}\right).
\ee
With this function ${\cal F}$ the supercharges \p{Z13} provide the supercharges for the $N=4$ supersymmetric $G_2$ model.

The superpotential $F(Z)$ entering the superfield action \p{Action} may be also easily restored to be
\be
F=\frac{1}{18}\left[ (1+4z)\log(1+4z)+(1-4z)\log(1-4z)\right]
\ee
for the case $g=f$. Here, the coordinate $z$ is related with $\phi$ as
\be \cos^2(3\phi) = 2z+\frac{1}{2}
\ee
and $\frac{m^2}{9}=72g$. When $g\neq f$ the superpotential reads
\bea
F&=& \frac{2\sqrt{\alpha}}{9a_1} \left[ (a_1 z +b_1 -\sqrt{\alpha})\log(a_1 z +b_1 -\sqrt{\alpha})-
(a_1 z +b_1 +\sqrt{\alpha})\log(a_1 z +b_1 +\sqrt{\alpha})\right]+ \nn \\
&& \frac{2\sqrt{\beta}}{9a_1} \left[ (a_1 z +b_1 +\sqrt{\beta})\log(a_1 z +b_1 +\sqrt{\beta})-
(a_1 z +b_1 -\sqrt{\beta})\log(a_1 z +b_1 -\sqrt{\beta})\right].
\eea
Here,
\be
g=\alpha \hat g,\; f=\beta \hat g, \qquad \frac{m^2}{9} = 72 {\hat g}
\ee
and 
\be 
\cos^2(3\phi) = a_1 z^2 +2 b_1 z +\frac{1}{2}
\ee
with
\be a_1 =\beta-\alpha,\quad b_1^2 =\frac{1}{2}(\alpha+\beta).
\ee
\subsection{$BC_2$ Calogero model}
Unlike the ordinary 3-particles Calogero model, the $BC_2, B_2,C_2$ and $D_2$ Calogero models are not translation invariant.
The actions for $BC_2, B_2,C_2$ Calogero models are all given by \cite{{bc2},{OP}}
\be\label{bc1}
S_{\rm BC_2}=\int dt \left \{ \frac{1}{2} {\dot y}{}_1^2 + \frac{1}{2} {\dot y}{}_2^2
-g_1 \left[ \frac{1}{(y_1-y_2)^2} + \frac{1}{(y_1+y_2)^2} \right]
- \frac{g_2}{2} \left[ \frac{1}{y_1^2} + \frac{1}{y_2^2}\right] \right \}.
\ee
When the coupling constant $g_2$ goes to zero, the action \p{bc1} degenerates to the action of $D_2$ Calogero model.

Similarly to the previously considered cases, after passing to the coordinates \p{coor2} the action \p{bc1} is simplified to 
\be\label{bc3}
S_{\rm BC_2}=\int dt\left \{ \frac{1}{2} e^{2u} \left( {\dot u}{}^2 +{\dot\phi}{}^2\right)
-2e^{-2u}\;\frac{g_1}{\cos^2(2\phi)}
-2e^{-2u}\;\frac{g_2}{\sin^2(2\phi)}\right \}.
\ee
Once again, comparing \p{bc3} with \p{Z14} one may find the function ${\cal F}$
\be\label{F2}
\frac{1}{72} m^2 \cF^2(\phi) = 2\left( \frac{g_1}{\cos^2(2\phi)}+\frac{g_2}{\sin^2(2\phi)}\right).
\ee
Clearly, the supercharges \p{Z13} with this function inserted, yield the Hamiltonian \p{Z14} which is just $N=4$ superconformal
Hamiltonian for the $BC_2, B_2,C_2$ and $D_2$ Calogero models.

Finally, the superpotential $F(Z)$ for this model reads
\be
F=\frac{1}{2}\left[ (1+4z)\log(1+4z)+(1-4z)\log(1-4z)\right]
\ee
for the case $g_1=g_2=g$. Here, the coordinate $z$ is related with $\phi$ as
\be \cos^2(2\phi) = 2z+\frac{1}{2}
\ee
and $\frac{m^2}{64}=g$. When $g_1\neq g_2$ the superpotential reads
\bea
F&=& \frac{\sqrt{\alpha}}{2a_1} \left[ (a_1 z +b_1 -\sqrt{\alpha})\log(a_1 z +b_1 -\sqrt{\alpha})-
(a_1 z +b_1 +\sqrt{\alpha})\log(a_1 z +b_1 +\sqrt{\alpha})\right]+ \nn \\
&& \frac{\sqrt{\beta}}{2a_1} \left[ (a_1 z +b_1 +\sqrt{\beta})\log(a_1 z +b_1 +\sqrt{\beta})-
(a_1 z +b_1 -\sqrt{\beta})\log(a_1 z +b_1 -\sqrt{\beta})\right].
\eea
Here,
\be
g_1=\alpha \hat g,\; g_2=\beta \hat g, \qquad \frac{m^2}{64} = {\hat g}
\ee
and 
\be 
\cos^2(2\phi) = a_1 z^2 +2 b_1 z +\frac{1}{2}
\ee
with
\be a_1 =\beta-\alpha,\quad b_1^2 =\frac{1}{2}(\alpha+\beta).
\ee

\subsection{Quantization}
In order to simplify the construction of the quantum version of the supercharges and the Hamiltonian, let us firstly pass to the new
coordinate
\be\label{x}
x=e^{u}.
\ee
Now, the classical supercharges read
\bea
&&
Q^i = \eta^i P_x + \frac{1}{x} \left[  \xi^i P_{\phi}
- \frac{m}{6}  \cF \bar \xi^i  - \frac{i}{2}  \frac{\cF^{\prime}}{\cF} \xi^2 \bar \xi^i
+ \frac{i}{2} \eta^2 \bar \eta^i - 2i  \xi^{(i} \bar \xi^{k)} \eta_k\right]\,, \nn\\
&&
\bar Q_i =   \bar \eta_i P_x + \frac{1}{x}\left[ \bar \xi_i P_{\phi}
+ \frac{m}{6}  \cF \xi_i  - \frac{i}{2}  \frac{\cF^{\prime}}{\cF} \xi_i \bar \xi^2
+ \frac{i}{2}  \eta_i \bar \eta^2 + 2i  \xi_{(i} \bar \xi_{k)} \bar \eta^k\right].
\label{q1}
\eea
With our choice of the Dirac brackets \p{Z10} and the $N=4$ superalgebra \p{sca}, we perform the quantization by replacing
the Dirac brackets by (anti)commutators using the rule\footnote{We decided to introduce the Planck constant $\hbar$ explicitly,
in order to keep  full control on  quantum corrections.}
\be
-i\left\{\; ,\; \right\}_{Dirac} = \hbar\left\{ \; , \; \right\},
\ee
and obtain the quantum algebra
\bea\label{scaq}
&&\left\{ Q^i , Q^j\right\}=0, \qquad \left\{ {\bar Q}_i , {\bar Q}_j\right\} =0, \nn \\
&&\left\{ Q^i , {\bar Q}_j\right\}=2 \hbar \delta^i_j  H_{quant} .
\eea 
In order to satisfy the quantum superalgebra \p{scaq} the supercharges get modified as
\bea
&&
Q^i = \eta^i P_x + \frac{1}{x} \left[  \xi^i P_{\phi}
- \frac{m}{6}  \cF \bar \xi^i  - \frac{i}{2}  \frac{\cF^{\prime}}{\cF} \xi^2 \bar \xi^i
+ \frac{i}{2} \eta^2 \bar \eta^i - 2i  \xi^{(i} \bar \xi^{k)} \eta_k+i\frac{\hbar}{2}\left(\eta^i -
\frac{\cF^{\prime}}{\cF}\xi^i\right)\right]\,, \nn\\
&&
\bar Q_i =   \bar \eta_i P_x + \frac{1}{x}\left[ \bar \xi_i P_{\phi}
+ \frac{m}{6}  \cF \xi_i  - \frac{i}{2}  \frac{\cF^{\prime}}{\cF} \xi_i \bar \xi^2
+ \frac{i}{2}  \eta_i \bar \eta^2 + 2i  \xi_{(i} \bar \xi_{k)} \bar \eta^k-i\frac{\hbar}{2}\left(\bar\eta_i -
\frac{\cF^{\prime}}{\cF}\bar\xi_i\right)\right].
\label{q2}
\eea
Let us stress that in \p{q2} the order of  all operators is strictly fixed so as to obey \p{scaq}.
The quantum Hamiltonian is also modified by  quantum corrections
\be
H_{quant}=H+H_q ,
\ee
where $H$ is the classical Hamiltonian \p{Z14} rewritten in the $x$-coordinate \p{x} and with all terms ordered
\bea
H \!&=&\! \frac{1}{2}  P^2_x +\frac{1}{x^2} \left[
\frac{1}{2}  P^2_{\phi}
- i  P_{\phi} \left (\eta^i \bar \xi_i - \xi^i \bar \eta_i \right )\right. \nn\\
\!&+&\! \frac{1}{72}  m^2 \cF^2
- \frac{i}{6}  m  \left ( \cF \eta^i  \xi_i
-\cF \bar \eta_i \bar \xi_i
- \frac{1}{2} \cF^{\prime} \xi^2 + \frac{1}{2} \cF^{\prime} \bar \xi^2 \right) \nn\\
\!&+&\! \frac{1}{4}  \left (\eta^2 \bar \eta^2 - \left(3 +\frac{\cF^{\prime}{}^2}{\cF^2}- 
\frac{\cF^{\prime \prime}}{\cF} \right)
\xi^2 \bar \xi^2 \right) \nn\\
\!&+&\! \left. \left ( \eta^i \bar \eta_i \xi^j \bar \xi_j - 2
\eta^i \xi^j \bar\xi_i \bar \eta_j \right) - \frac{1}{2}
\frac{\cF^{\prime}}{\cF} \left(\eta^i \xi_i \bar \xi^2 +
\xi^2  \bar \eta_i \bar \xi^i \right) \right],\label{q3} 
\eea
and the quantum corrections read
\be
H_q = -\frac{\hbar}{2 x^2}\left[\eta^i\bar\eta_i -\frac{\cF'}{\cF}\left( \eta^i \bar\xi_i+\xi^i\bar\eta_i\right) -
\left( 3+\frac{\cF'{}^2}{\cF^2}-\frac{\cF''}{\cF}\right) \xi^i \bar\xi_i \right]+ 
 \frac{\hbar^2}{8x^2}\left( 3- \frac{\cF'{}^2}{\cF^2}+2\frac{\cF''}{\cF}\right).
\ee
These expressions provide the description of the general $N=4$ superconformally invariant 3-particles mechanics (with
the translational invariance) in the basis with decoupled center of mass.

\section{Conclusion}
In this paper we have constructed the most general $N=4$ superconformal 3-particles systems with translation invariance.
In the basis with decoupled center of mass the supercharges and Hamiltonian possess one arbitrary function which
defines  all potential terms. We have shown that with the proper choice of this function one may describe the
standard, $A_2$ Calogero model with $N=4$ supersymmetry, as well as $BC_2, B_2,C_2$ and $D_2$ Calogero models.
The main property of all these systems is that even with the coupling constant equal to zero they still contain
nontrivial interactions in the fermionic sector. In other words there are infinitely many inequivalent
$N=4$ supersymmetrizations of the free action depending upon one arbitrary function. Just this property makes all construction
rather nontrivial. Indeed, one cannot start from the supercharges for the free model and then try to find the coupling-constant
dependent terms to get, for example, $N=4$ supersymmetric Calogero model. The freedom in the "free" supercharge has to be
fixed firstly in a proper way. This is the main difference between $N=2$ and $N=4$ supersymmetric models.
We also considered the quantization and explicitly showed how the supercharges and Hamiltonian get modified. The bosonic core of the
quantum Hamiltonian contains  $\hbar^2$-terms in  full agreement with the results presented in \cite{GLP}. This is again a novel
feature of $N=4$ supersymmetry which appeared already in the one particle case \cite{IKP}.

It is worth noting that our construction is closely related with the considerations in \cite{{IKP},{Tolik}}.

An immediately interesting issue for future study is provided by the cases with $n \geq4 $ particles. The corresponding superfield construction
is not too complicated. But some new features appear in these cases.
First of all, in the superfield Lagrangian there may appear some additional terms. So, the most general action is not of the
form \p{Action} with the superfunction $F$ depending on $(n-2)$ scalars. In addition, there is  freedom to choose
as the basic superfields two types of $N=4$ superfields with one physical boson: the standard one and the "mirror" supermultiplet
(see e.g. \cite{ID} and references therein). Finally, the following serious problem arises.
Throughout the paper we worked in the rather specific coordinate system in which the $N=4$ supersymmerization has the simplest
form. In this system (with the center of mass decoupled) we also select one field $u$ to be a proper dilaton field. The arbitrary
function which defines the potential terms and self-interaction of the fermionic coordinates depends on the rest coordinates --
one scalar field in the case of the 3-particles system. In the cases with $n\geq 4$ particles the main problem is to specify the
arbitrary function in order to have 1) flat bosonic kinetic terms, 2) proper potential terms in the bosonic sector. This is not so easy
and we are planning to consider all these tasks elsewhere. 

\section*{Acknowledgements}

We are grateful to A.~Galajinsky, O.~Lechtenfeld and A.~Nersessian for valuable discussions.
\vskip .2cm
S.K. and A.S. thank Laboratori Nazionali di Frascati, INFN, where this work was completed for kind hospitality
and financial support.  This work was partially supported by the European Community Human Potential
Program under contract MRTN-CT-2004-005104 \textit{``Constituents, fundamental forces and symmetries of the universe''}, 
by  INTAS under contract 05-7928 and by grants RFBR-06-02-16684, 06-01-00627-a and  DFG~436 Rus~113/669/03.

\end{document}